\documentclass[superscriptaddress,twocolumn,showpacs,preprintnumbers,amsmath,amssymb]{revtex4}
\usepackage{graphicx}
\usepackage{epsfig}
\usepackage{dcolumn}
\usepackage{bm}
\usepackage[latin1]{inputenc}
\newcommand{\chem}[1]{\ensuremath{\mathrm{#1}}}
\newcommand{\un}[1]{\ensuremath{\unskip\,\mathrm{#1}}}

\begin{document}
\title {Spontaneous spiking in an autaptic Hodgkin-Huxley set up}
\author{Yunyun Li}
\affiliation{Institut f\"ur Physik, Universit\"at Augsburg, Universit\"atsstr. 1, 86159 Augsburg, Germany}
\author{Gerhard Schmid}
\affiliation{Institut f\"ur Physik, Universit\"at Augsburg, Universit\"atsstr. 1, 86159 Augsburg, Germany}
\author{Peter H\"anggi}
\affiliation{Institut f\"ur Physik, Universit\"at Augsburg, Universit\"atsstr. 1, 86159 Augsburg, Germany}
\author{Lutz Schimansky-Geier}
\affiliation{Institut f\"ur Physik, Humboldt Universit\"at zu Berlin,
  Newtonstr. 15,  12489 Berlin, Germany}
\date{\today}

\begin{abstract}
  The effect of intrinsic channel noise is investigated for the dynamic
  response of a neuronal cell with a delayed feedback loop. The loop is
  based on  the so-called autapse phenomenon in which dendrites
  establish not only connections to neighboring cells but as well to its own
  axon.  The biophysical modeling is
  achieved in terms of a stochastic Hodgkin-Huxley model containing such a built in delayed feedback.
  The fluctuations stem from intrinsic channel noise, being
  caused by the stochastic nature of the gating dynamics of ion channels. The influence
  of the delayed stimulus is systematically analyzed with respect to
  the coupling parameter and the delay time in terms of the  interspike
  interval histograms and the average interspike interval. The delayed
  feedback manifests itself in the occurrence of bursting and a rich
  multimodal interspike interval distribution, exhibiting a delay-induced
  reduction of the spontaneous spiking activity at characteristic
  frequencies. Moreover, a specific frequency-locking mechanism is detected for the mean
  interspike interval.
\end{abstract}
\maketitle

\section{Introduction}
\label{sec:intro} Time-delayed feedback is a common mechanism
relevant in many biological system including excitable gene
regulatory circuits \cite{cc1} and human balance \cite{cc2}. It has
been investigated not only in  biological systems, but also in a
wider area: This includes but is not limited to semiconductor
superlattices \cite{sh}, chemical oscillators (CO oxidation on
platinum) \cite{th} and photosensitive Oregonator models \cite{gao}.
These types of systems have been formulated mathematically in terms
of  recurrent models \cite{cri-1,cri}. Recurrent connections
introduce delayed feedback loops which may dramatically change the
dynamic behavior of the system. It is  known that delayed feedback
presents an efficient method to control chaos or turbulence via
stabilizing the unstable periodic orbits (UPOs) embedded in the
chaotic attractor \cite{gao1}. Feedback can be used to stabilize
periodic orbits \cite{th}, or to control the coherence resonance
\cite{sh, sch, ku2}; the latter has also been studied experimentally
for bistable system with delayed feedback \cite{ex}.

Over the past decades, neurobiologists found out that axons
propagate the neuron's electrical signal to other neurons, and may
sometimes feedback to the same neuron's dendrites~\cite{aut-new1,
  aut-new2, aut-new3, aut-new4}.
These auto-synapses which establish a time-delayed feedback
mechanism on a cellular level, are called autapses and were
described by Van der Loos and Glaser in 1972 \cite{Loos1972}. Since
the discovery of such autapse in pyramidal cells in the cerebral
neocortex, they were found in about 80\% of all analyzed neurons
including neurons of the human brain \cite{aut3}. However, what
functional significance they offer in the neural systems is still
not fully understood, as the autapses exhibit a broad range of delay
time scales: these range from a few milliseconds to tenths of
milliseconds \cite{cri}.

It was established that delayed feedback can induce bursting
\cite{ku2}. Its role for coding and processing of information in the
brain has been evidenced experimentally for a variety of neural
systems. However, the mechanisms leading to bursting are still open
to debate; in particular the presence of noise is expected to play a
key role.

Within this work we aim to contribute to this objective by
considering the influence of intrinsic noise. It has been
demonstrated, that noise leads to nontrivial effects in neuronal
dynamics \cite{Lindner}. Some typical examples are  stochastic
resonance phenomena \cite{Gammaitoni,hanggi,Schmid_EPL, Shuai_EPL},
coherence resonance \cite{meinhold,schi1,schi2,bur} and dynamical
synchronization phenomena \cite{sy,sy0,sy1,sy2}. In our situation
there occurs an intrinsic source of noise which is due to the
stochastic gating of ion channels; i.e. so-called channel noise. The
latter is inherently coupled to the properties of the axonal cell
membrane and, {\itshape a priori}, cannot be neglected \cite{noi1}.
Interestingly, this intrinsic noise effects the neuronal signaling
at different levels: It was shown that it may control the occurrence
of spontaneous action potentials, can improve the output of signal
quality \cite{Schmid_EPL} and may also account for the reliability
of propagation \cite{Ochab}, to name but a few.

The manuscript is organized as follows: in Sec. \ref{sec:model} we
present the biophysical model:  It is given in terms of a stochastic Hodgkin-Huxley
model containing a time-delayed feedback current.  In Sec.~\ref{sec:detdyn} we
discuss the repetitive firing that occurs for the deterministic dynamics (i.e. in absence of noise)
of this Hodgkin-Huxley model with delayed feedback. The stochastic
dynamics are addressed in Sect.~\ref{sec:stochdyn}. Finally, we present our conclusions with
 Sec.~\ref{sec:conclusion}.

\section{Model set up}
\label{sec:model} In 1952, Hodgkin and Huxley proposed an
archetypical model for cell excitability and signal transmission
along the axon of a neuronal cell \cite{hh}. They  postulated a set
of continuous parallel pathways for the passage of the ionic and
capacitive currents to represent the electrical properties of the
neuronal cell membrane. The Hodgkin-Huxley model is widely regarded
as a milestone achievement in biophysics and especially in
electrophysiology. Its applicability was extended to more complex
systems beyond the  originally studied excitability of the squid
giant axon. Experimental evidence of the channel noise has resulted
in stochastic generalizations of the Hodgkin-Huxley model by
considering the stochastic dynamics of the ion channel gating
\cite{white}. These generalizations allow for microscopic modeling
of the occurrence of spontaneous spiking and the phenomenon of
biological Stochastic Resonance \cite{hanggi}. In this context, it
was shown recently, that channel noise improves the reliability for
signal transmission along neuronal axons \cite{Ochab}.

\begin{figure}
  \centering
  \includegraphics[width=0.5\textwidth]{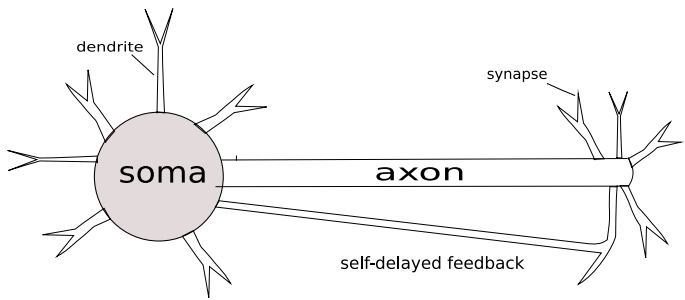}
  \caption{Sketch of an autapse: Sketch of a neuronal cell exhibiting
    a self delayed feedback mechanism.}
  \label{fig:autapse}
\end{figure}
We mathematically model the delayed feedback induced by an autapse
(sketched in Fig.~\ref{fig:autapse})
 by a feedback mechanism within a Hodgkin-Huxley model of the type put forward by Pyragas
 \cite{del}. By an additional pathway in the dynamics of
the membrane potential we consider the autapse leading to a delayed
stimulus $I_\mathrm{\tau}(t)$. Particularly, we consider an
excitable dynamics for the membrane potential $V(t)$, reading:

\begin{align}
  \label{eq:comp_mod}
  C \frac{\mathrm{d} V(t)}{\mathrm{d} t} = & -n^4(t)
  g_{\chem{K}}^{\mathrm{max}} \left(V(t)-V_{\chem{K}}\right) -
  g_{\chem{L}}\left(V(t)-V_{\chem{L}}\right) - \nonumber \\\\&
  \hspace{0.cm}-m^3(t)h(t)
  g_{\chem{Na}}^{\mathrm{max}}\left(V(t)-V_{\chem{Na}}\right) +
  I_\mathrm{\tau}(t)\,  \nonumber
\end{align}
wherein the autaptic delayed stimulus
\begin{align}
\label{curr}
  I_\mathrm{\tau}(t) = & \epsilon \left[V(t-\tau)-V(t)\right]\,,
\end{align}
is proportional to the difference of the membrane potential at time
$t$ and that at an earlier time $t-\tau$. Here, $\epsilon$
corresponds to the coupling strength of the autapse mechanism and
the finite delay time $\tau$ refers to the specific time delay
caused by the autapse which is due to a finite signal propagation
speed. The delay time $\tau$ is representing the elapsed time
associated with the axonal propagation prior to the signal recurring
onto the neuron; $V(t-\tau)$ is the membrane potential at the
earlier time t-$\tau$.

The autaptic delayed stimulus in Eq.~\ref{curr} results in an
excitatory coupling mechanism in which
spiking of a cell at an
earlier time $t-\tau$ favors the initiation of a spiking event of the same cell
at time $t$. We note, that the ansatz for the  delayed
self-stimulus  corresponds to electrotonic
interaction, i.e. we consider an idealized situation wherein the autaptic delayed stimulus
is proportional to the difference of the pre- and post-synaptic membrane
potentials. Moreover, as autapses are typically formed by chemical synapses, our
modeling simulates the complex biophysical temporal evolution of the synaptic
conductance by invoking a constant coupling strength
$\epsilon$.

In Eq.~\eqref{eq:comp_mod}, $C$ denotes the
membrane capacitance, $V(t)$ the time-dependent membrane potential,
$V_\chem{L}$ the leakage potential, $V_\chem{K}$ and $V_\chem{Na}$
the reversal potentials for the potassium and sodium currents. The
leakage conductance is given by $g_{\chem{K}}$, the potassium and
sodium maximum conductances read $g_{\chem{K}}^{\mathrm{max}}$ and
$g_{\chem{Na}}^{\mathrm{max}}$, respectively. The functions  $ m(t)
$, $ n(t) $ and $ h(t) $ in Eq. (\ref{eq:comp_mod}) denote the
so-called stochastic gating variables (see below), describing the
mean fraction of open gates of the sodium and potassium channels at
time $t$. The typical values of parameters in our numerical
simulation are those of the original Hodgkin-Huxley model \cite{hh};
i.e., $C$=1 $\mathrm{\mu F/cm^2}$, $g_\chem{K}^{\mathrm{max}}=120
\mathrm{mS/cm^2}$, $g_\chem{Na}^{\mathrm{max}}=36 \mathrm{mS/cm^2}$,
$g_\chem{L}=0.3 \mathrm{mS/cm^2}$, $V_\chem{K}=-77 \mathrm{mV}$,
$V_\chem{Na}=50 \mathrm{mV}$ and $V_\chem{L}=54.4\mathrm{mV}$.

The stochastic dynamics of the gating variables $m(t)$, $n(t)$ and $h(t)$
depend on the voltage-dependent opening and closing rates $\alpha_i(V)$
and $\beta_i(V)$ ($i=m, h, n$), which read \cite{hh}:
\begin{subequations}
\label{eq:gating}
\begin{align}
  \label{eq:rates-m}
  \alpha_{m}(V) &=\frac{0.1(V+40)}{1 - \exp\left\{-\, (V+40)/10
    \right\}}\, , \\
  \beta_{m}(V)  &= 4\, \exp\left\{ -\, (V+65)/18 \right\}\, , \\
  \alpha_{h}(V) &= 0.07  \, \exp\left\{ -\, (V+65)/20 \right\}\, , \\
  \beta_{h}(V)  &= \frac{1}{1 + \exp\left\{-\, (V+35)/10 \right\} }\, ,  \\
  \alpha_{n}(V) &= \frac{0.01(V + 55)}{1 - \exp \left\{-\, (V + 55)
      / 10 \right\}}\, ,\\
  \label{eq:rates-n}
  \beta_{n}(V)  &= 0.125 \, \exp\left\{ -\, (V+65)/80 \right\}\, .
\end{align}
\end{subequations}

The resulting stochastic dynamics is then modeled  by a
Fokker-Plank-type dynamics for the individual gates. Specifically,
this  dynamics emerges within a large system size-expansion of the
underlying Markovian master-equation dynamics for the dynamics of
the number of open gates \cite{noi1,tuckwell}. This so obtained
Langevin-dynamics is then interpreted in the It$\hat{o}$ sense,
reading explicitly:
\begin{equation}
  \label{eq:correlator-a}
\frac{\mathrm{d} i}{\mathrm{d t}}=\alpha_i(V)(1-i)-\beta_i(V)i+\xi_i(t),
\end{equation}

with $i=m, h, n$. Here, $\xi_i(t)$ ($i=m, h, n$) denote independent
Gaussian white noise sources  with vanishing mean and auto-correlation
functions:
\begin{subequations}
\label{eq:noisecor}
\begin{align}
\langle \xi_m(t)\xi_m(t') \rangle&=\frac{(1-m)\alpha_m+m \beta_m}{N_{Na}}\delta(t-t'),\\
\langle \xi_h(t)\xi_h(t') \rangle&=\frac{(1-h)\alpha_h+h\beta_h}{N_{Na}}\delta(t-t'),\\
\langle \xi_n(t)\xi_n(t') \rangle&=\frac{(1-n)\alpha_n+n\beta_n}{N_K}\delta(t-t').
\end{align}
\end{subequations}

Note, that the noise strengths are determined by the number of
potassium and sodium channels $N_\chem{K}$ and $N_\chem{Na}$.
Throughout this work we assume  constant ion channel densities; i.e.
following the original Hodgkin-Huxley model we use  $18$ for
potassium and $60$ for the sodium channels per $\mu \mathrm{m}^{2}$
\cite{hh}. An integration step $\Delta t=0.001 \un{ms}$ was used in
the simulations and for the generation of the Gaussian distributed
random numbers, the Box-Muller algorithm \cite{bm} was used.

\section{Deterministic dynamics}
\label{sec:detdyn}

In the absence of both, the delayed feedback, i.e. $\epsilon = 0$,
and zero noise strength, with the latter being formally achieved by
$N_\chem{Na} \to \infty$ and $N_\chem{K} \to \infty$, the original
Hodgkin-Huxley dynamics are recovered. These exhibit a single stable
fixed point, which is the {\itshape rest state} with the rest
potential $V_\mathrm{rest} \approx -65 \un{mV}$.  Any temporary disturbance,
e.g. caused by an applied external current stimulus, decays and the
system relaxes back to the rest state. Note that upon a proper
choice of the initial condition, a single action potential may
occur, before the system relaxes to the  rest state \cite{textbook}.

\begin{figure}
  \begin{center}
    \includegraphics{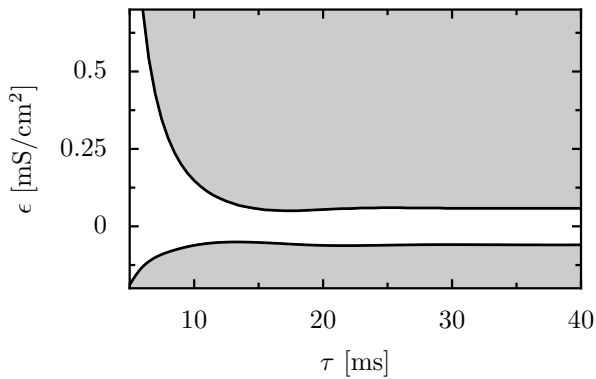}\\
    \caption[]{\label{fig:1} The phase diagram $\epsilon - \tau$ for repetitive firing
      within the Hodgkin Huxley model containing the delay coupling strength $\epsilon$ is
      depicted {\it vs.} the delay time $\tau$.  The two black solid lines indicate the boundaries  of
      critical coupling strength $\epsilon$ versus the delay
      time $\tau$ at which the behavior changes from the rest state to a repetitive firing.
      The  gray areas give the parameter regime for which
      repetitive firing is observed.  }
  \end{center}
\end{figure}

In presence of the autaptic delay coupling, the fixed point solution
remains stable, but upon increasing the delay time parameter $\tau$,
there emerges now a  stable, oscillatory solution, presenting
repetitive (tonic) spiking. To investigate this repetitive spiking
behavior in more detail, we systematically varied the two parameters
($\epsilon$ and $\tau$) of the delay coupling scheme. In doing so,
we determined the first occurrence of repetitive firing after a
spike was created. The resulting phase diagram with the positive and
negative critical coupling strength $\epsilon_{c}^{+}(\tau)$ and
$\epsilon_{c}^{-}(\tau)$, respectively, is depicted in Fig.
\ref{fig:1}. We point out the resulting asymmetry between positive
and negative coupling $\epsilon$.

For sub-threshold coupling strength, i.e. $ \epsilon_{c}^{-}( \tau )
< \epsilon < \epsilon_{c}^{+}(\tau)$ (in the white regions of Fig.
\ref{fig:1}), excitations die out and the system relaxes to the rest
state. For supra-threshold coupling, i.e. $\epsilon >
\epsilon_{c}^{+}(\tau)$ or $\epsilon < \epsilon_{c}^{-}(\tau)$ a
repetitive firing is observed. If the delay time becomes shorter
than the {\itshape refractory time interval},  being ca.
$12\un{ms}$, which is the time the system needs for spiking and
returning to the rest state, the modulus of the critical coupling
strength of the bifurcation increases drastically.  This is due to
the fact, that in the so-called {\itshape undershoot phase}
(i.e. the part of the course of the action potential when the membrane potential is smaller than the membrane potential at rest)  the
neuron is rather insensitive to any stimuli. Hence, a stronger
coupling strength is needed in order to excite the system from the
{\itshape undershoot phase} and to obtain repetitive firing. In case
of larger delay the value of the critical coupling saturates. For
longer delay times more than one spike may fit into time interval
given by $\tau$. Consequently, doublets, triplets and multiplets may
appear.

Below, we shall present examples where we have chosen $\tau
=35$\un{ms} and used a positive coupling strength.  The critical
value for repetitive firing corresponds to $\epsilon_c^{+}=
0.059\un{mS/cm^2}$. For larger coupling $\epsilon >
\epsilon_{c}^{+}(\tau)$ an action potential induces repetitive
firing. In section \ref{sec:stochdyn} we refer to this regime as
{\it supra-threshold delay coupling regime}.  In case of lower
coupling strength; i.e.  $\epsilon < \epsilon_{c}^{+}(\tau)$ the state
of a repetitive firing is excitable and noise is needed to
support the repetition of an action potential. In the next section
\ref{sec:stochdyn} this regime will be addressed as {\it
sub-threshold delay coupling regime}. Remarkably, for negative
coupling $\epsilon < 0$ one finds an overall similar similar
behavior that differs, however, quantitatively (not shown).

\section{Stochastic dynamics of the Hodgkin-Huxley model with delayed feedback}
\label{sec:stochdyn}

\begin{figure}
  \begin{center}
    \epsfig{figure=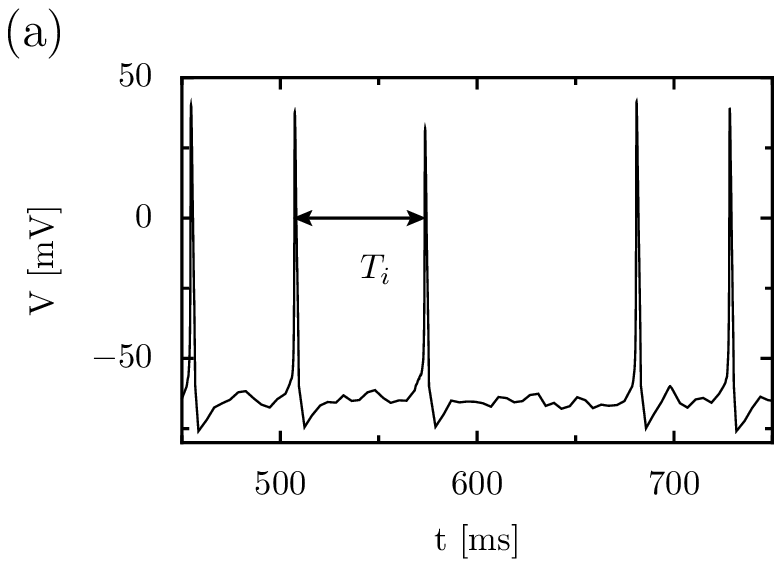,scale=1}\\
    \epsfig{figure=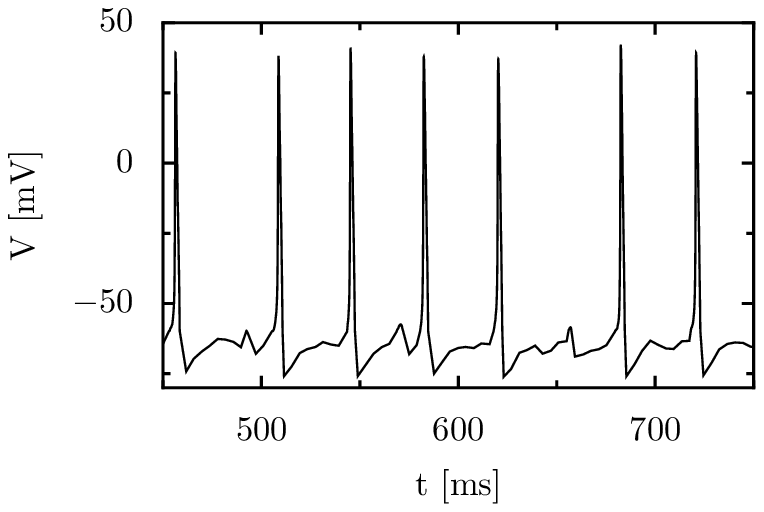,scale=1}\\
    \caption[]{\label{fig:spiketrain} Simulated spike-trains: (a) Membrane potential $V(t)$ from the
      stochastic Hodgkin-Huxley model in absence of feedback, i.e. with  $\epsilon=0$ versus time $t$; (b)
    $V(t)$ versus  time $t$ in presence of finite feedback of strength  $\epsilon =0.07
      \mathrm{mS/cm^2}$ exhibiting repetitive spiking for a chosen delay time
      $\tau=35$\un{ms}. The number of potassium ion channels is set
      to $N_{\chem K} =300$ and those of the sodium ion channels to
      $N_{\chem{Na}}=1000$. The spontaneous, i.e.  noise induced,
      action potentials  exhibit a bursting-like behavior in case of
      the delay coupling (b).
      In (a) we indicated with the double-arrow
      line a typical occurring noise-induced interspike interval
      $T_i$.}
  \end{center}
\end{figure}

In Fig.~\ref{fig:spiketrain} the simulated membrane potential $V(t)$
is shown for an exemplarily chosen noise level and for two  cases,
namely in panel (a)  without delay coupling ($\epsilon=0$) and in
panel (b) with a finite delay coupling $(\epsilon > 0$). In absence
of delay coupling the occurrence of spontaneous, i.e. noise induced
actions potentials occur  irregular while a characteristic bursting
pattern can be detected in presence of finite delay-coupling
$\epsilon$.

In order to explain the spiking dynamics quantitatively, we
determined the time intervals between two spike events $T_i$, see in
Fig.~\ref{fig:spiketrain}(a), as obtained from simulations of the
stochastic Hodgkin-Huxley model with delayed feedback coupling, cf.
Eq.~\eqref{eq:comp_mod}. Note that in order to detect the
occurrences of action potentials from the simulated membrane
potential dynamics $V(t)$, we  did define a specific threshold
barrier for detection. In particular, whenever the membrane
potential $V(t)$ exceeds the value of $0\un{mV}$, the occurrence of
an action potential is assigned. In fact, the so determined spike
occurrences depend only weakly on the actual choice of the detection
barrier \cite{Schmid_EPL}.

These observed stochastically emerging interspike intervals $T_i$
the render the (normalized) interspike interval histograms (ISIH) and the mean
interspike interval $\langle T\rangle$
\begin{align}
\label{eq:def_meantime}
\langle T\rangle=\frac{1}{N}\sum\limits_{i=1}^N T_i\, ,
\end{align}
where $N$ denotes the number of spikes obtained in the individual
numerical simulation. The bin width for the histograms has been set at $ 0.2 \un{ms}$.
Both these quantities form the basis of our
analysis of spike trains.

We begin this study with the dynamics of zero delay coupling for the
neuron, i.e. with $\epsilon=0$.  The distribution of the interspike
intervals at a fixed channel noise intensity is depicted in
Fig.~\ref{fig:isihdifferente}.  For large time intervals, the
distribution decays exponentially. Small time intervals are
suppressed because of the neuron's refractory state.
For $\epsilon=0$ the ISIH
exhibits one broad maximum located around the internal time scale
${\cal T}_{\mathrm{int}}$ .
This time scale systematically diminishes with increasing noise
level \cite{Schmid_EPL,Lindner}. The minimal time scale for
undistorted spikes is around the refractory time of
$\approx 12 \un{ms}$.

In presence of noise and finite feedback coupling, the neuron still
can return to its fix-point with nonvanishing probability. This is
why the peak at the intrinsic time scale and a broad exponential
decay is still detected in the ISIH in
Fig.~\ref{fig:isihdifferente}. 
However, the ISIHs become multimodal in case of finite, nonvanishing
feedback. Several maxima occur at larger time intervals and, in
between, a number of deep dips are observed.  Such forms of the
distributions of the ISIH are  indicators of an enhanced coherence
\cite{sy0,Lindner}. Whereas in the sub-threshold regime (dashed
line) the additional extrema merely show up, they become
increasingly dominant in the case of supra-threshold coupling (note
the logarithmic scale of the ordinate in
Fig.~\ref{fig:isihdifferente}).

\subsection{Sub-threshold delay coupling: $\epsilon <
  \epsilon_{c}^{+}(\tau)$}

\begin{figure}
\begin{center}
\includegraphics{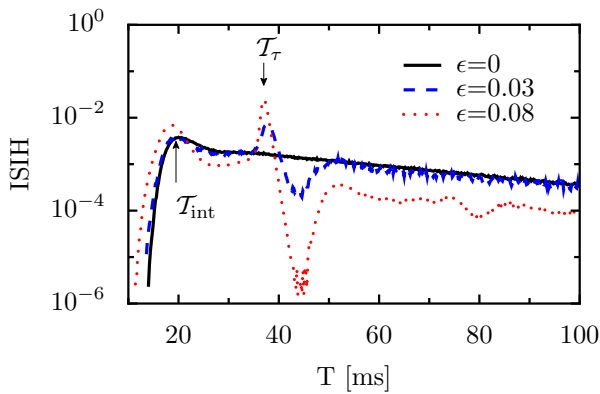}\\
\caption[]{\label{fig:isihdifferente} (Color online) Interspike
  interval histograms (ISIH) versus the interspike interval $T$ of the stochastic Hodgkin-Huxley model with a
  delay coupling mechanism at different coupling strengths $\epsilon$, as indicated in the
  figure in  units of  \un{mS/cm^2}.  The bin width  for $T$ has been chosen throughout this work  at $ 0.2 \un{ms}$. The chosen number of potassium ion channels is
  $N_\chem{K} = 150$, while the number of sodium ion channels is set at $N_\chem{Na} = 500$.
  The delay time $\tau$ is set at  $35$\un{ms}.}
\end{center}
\end{figure}

In the case of sub-threshold coupling the repetitive oscillatory
spiking is  noise supported, as a non-vanishing delayed stimulus
$I_\mathrm{\tau}(t)$ shifts the stable fix-point towards the
threshold for excitation.
Consequently, the generation of spikes becomes more likely. However,
the dynamics is still excitable; i.e. a threshold still
exists and noise is needed to induce its excitation.
Consequently, a  sub-threshold delayed stimulus favors some characteristic interspike interval time  ${\cal
  T_{\tau}}$. As far as the system remains
excitable, an activation time ${\cal T}_\mathrm{act}$ after the
delayed stimulus did set in  is necessary in order to create the
next noise-induced spike. Therefore, the time  ${\cal T}_{\tau}$ is not
equal to the delay time $\tau$, but equals the sum of the delay time
and an activation time ${\cal T}_\mathrm{act}$, being around
$2\un{ms}$ in this considered case.

\begin{figure}
\begin{center}
\includegraphics{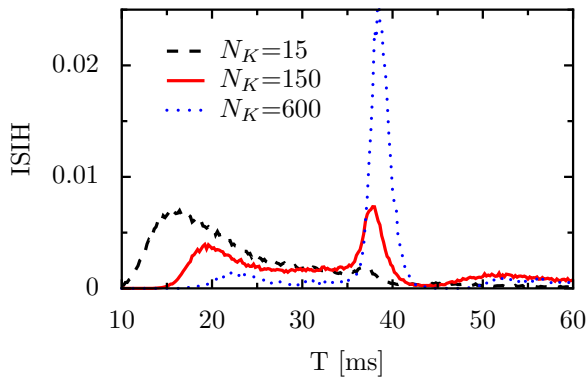}\\
\caption[]{\label{fig:isihnoise} (Color online) Interspike interval
  histogram (ISIH) {\it vs.} interspike duration $T$ for different noise levels corresponding to different
  number of sodium and potassium channels as indicated. The chosen delay time
  is : $\tau=35 \un{ms}$; the coupling strength is set at $\epsilon=0.03
  \un{mS/cm}^{2} <\epsilon_{c}^{+}(\tau)$ and $N_\chem{Na} = \frac{10}{3} N_\chem{K}$.}
\end{center}
\end{figure}
In addition, there occurs a suppression of certain time scales when
the delay coupling is applied to the neuronal dynamics, cf.
Fig.~\ref{fig:isihdifferente}.  For positive $\epsilon$ the system
favors frequencies associated to the time ${\cal T}_{\tau}$ and
suppresses frequencies slightly larger than those. The feedback
hinders the generation of spikes by noise if the feedback current
$I_\mathrm{\tau}(t)$ approaches values around the polarized
refractory period.

In the sub-threshold regime the intrinsic time scale ${\cal
T}_{\mathrm{int}}$ and correspondingly the height and the position
of the associated peak distinctly depends  on the noise level. In
Fig.~\ref{fig:isihnoise} we depict results from simulations for
different noise levels, i.e. different numbers of embedded ion
channels. With increasing noise level the generation of spontaneous
spikes by noise gains  influence substantially. As a result, the
intrinsic time scale becomes changed and the larger time intervals
reflect the broad, exponentially decaying ISIH. The former sharply
peaked maximum at ${\cal T}_{\tau}$ switches towards a noise supported
broader peak at smaller times, and the maximum shifts with growing
noise towards the minimal time for a interspike interval located at
the refractory time.
Interspike intervals around this minimal time are promoted mainly by
the presence of noise. They do not necessarily relate to the delay
mechanism. In contrast, the time scale ${\cal T}_{\tau}$ induced by the
delayed feedback is only slightly affected by the noise level. This
is  a somewhat striking feature recalling the fact that  in the case
of an excitable dynamics  these spikes in fact require the presence
of noise to become activated. The robustness of the positions of the
time scales as well as their separation from the noisy time scale in
terms of the resulting bimodal shape of the ISIH thus indicates a
noisy synchronization phenomenon \cite{sy1,sy2,sch,ku}.

\begin{figure}
\begin{center}
\includegraphics{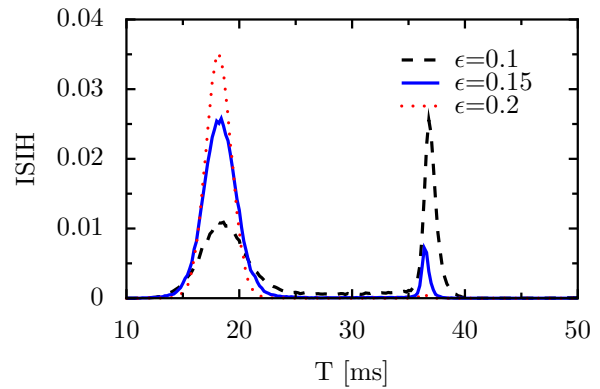}
\caption[]{\label{fig:isihe} (Color online) Interspike interval
  histogram (ISIH) for different values of the supra-threshold coupling
  strength $\epsilon > \epsilon_{c}^{+}(\tau)$: The distribution of the
  interspike intervals is depicted for fixed delay time $\tau=35
  \un{ms}$, the numbers of ion channels are $N_\chem{K}=150$ and
  $N_{\chem{Na}} = 500$. The values of the coupling strengths
  indicated in the plot are in units of  \un{mS/cm^2}.}
\end{center}
\end{figure}

\subsection{Supra-threshold delay coupling: $\epsilon > \epsilon_{c}^{+}(\tau)$}

We next consider the case with coupling strengths beyond
the  critical strength, i.e., $\epsilon > \epsilon_{c}^{+}(\tau)$.
Here, the spikes repeat deterministically.
However, noise-induced skipping of spikes \cite{sch-a,LNP} leads to a
transition from the oscillatory state with repetitive firing to the
excitable state. In turn, the {\itshape back-transitions} from the
excitable to the oscillatory state are caused by noise-induced spontaneous
spikes, cf. Fig.~\ref{fig:spiketrain}. The spikes repeated
deterministically lead to sharp peaks in the ISIH, while the
noise-induced {\it back-transitions} from excitable to the oscillatory
state result in a broad distribution with
exponentially decaying tail. In
Fig.~\ref{fig:isihe}, the resulting ISIH is depicted for various
supra-threshold coupling strengths.
Upon the chosen parameter
values for the noise level, i.e. the number of sodium and potassium
ion channels, and the coupling parameters $\epsilon$ and $\tau$, the ISIH therefore
exhibits sharp peaks and more or less pronounced broad background:
For stronger coupling or weaker noise the broad background
diminishes. Contrarily, the peak height at ${\cal T}_{\tau}$ shows a strong dependence on
the noise level, indicating the competitive interplay between
channel noise and the delayed feedback mechanism: With increasing
noise level the height of
the peak at ${\cal  T_{\tau}}$ decreases, while the intrinsic time scale acquires
increasing influence (depicted for the
sub-threshold coupling regime in Fig.~\ref{fig:isihnoise} ).
Note that we have selected the delay time
($\tau=35\un{ms}$) so that the separation between the broad background
which is peaked at the intrinsic time scale $\cal T_{\mathrm{int}}$ and the delayed
induced sharp peaks at ${\cal T}_{\tau}$ and multiples thereof is
educible and visible with the ISIHs, cf. Fig.~\ref{fig:isihnoise} and
Fig.~\ref{fig:isihe}.

Strikingly, with increasing
coupling strength the multimodal structure collapses to a unimodal
one, cf Fig.~\ref{fig:isihe}.
The most probable interspike interval now centers at one value.
Due to this supra-threshold driving, each spontaneous spiking event
is repeated periodically and even the noise dominated scale now
collapses towards a narrow peak. Note also that the distribution
gains in sharpness  as the coupling strength increases, cf.
Fig.~\ref{fig:isihe}.

\begin{figure}
\begin{center}
\includegraphics{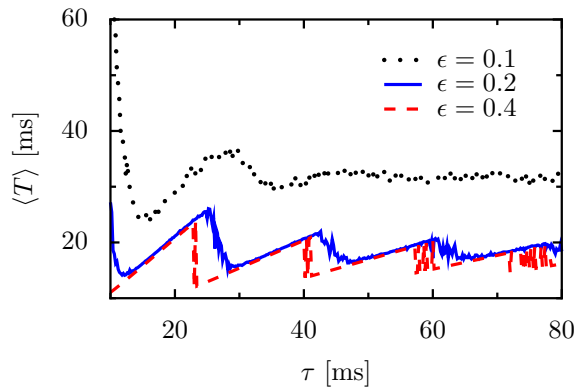}
\caption{(Color online) Average interspike interval $\langle
T\rangle$ as function of the delay time $\tau$: The mean interspike
interval $\langle
  T\rangle$, cf. Eq.~\eqref{eq:def_meantime}, is plotted versus
  the delay time $\tau$ for the following parameters: the used ion
  channel numbers are
  $N_\chem{K}=150$, $N_\chem{Na}=500$ and the coupling strengths are set at $\epsilon=0.1 \mathrm{mS/cm^2}$
  (black dashed line), $\epsilon=0.2 \mathrm{mS/cm^2}$ (red dotted
  line), and
  $\epsilon=0.4 \mathrm{mS/cm^2}$ (blue solid line).}
\label{fig:mean}
\end{center}
\end{figure}

\begin{figure}
\begin{center}
\includegraphics{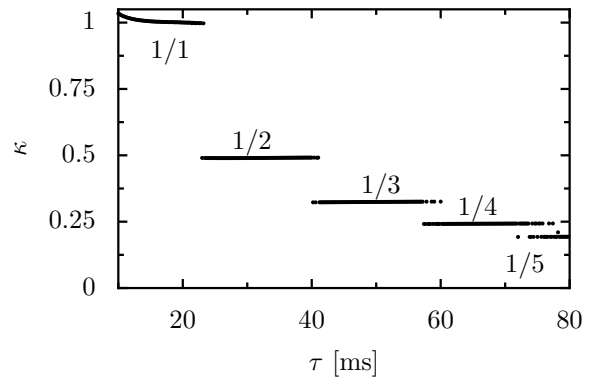}
\caption{\label{fig:kappa} The inverse of the number of spikes that
fit into the delay-time interval $\kappa$ is plotted as function of
the delay time $\tau$ for $N_\chem{K}=150$, $N_\chem{K}=500$ and a
delay coupling strength of $\epsilon=0.4 \mathrm{mS/cm^2}$.}
\end{center}
\end{figure}

Let us focus  on the dependence of the interspike intervals on the delay time $\tau$.
For this purpose we consider  the mean interspike interval, cf.
Eq.~\eqref{eq:def_meantime}. This quantity inherits the
characteristic dependence on $\tau$ since the ISIH shape is
unimodal. Larger deviations from the mean interspike interval are
found during transitions between different synchronized states where
the histogram becomes multimodal.

In Fig.~\ref{fig:mean} the mean interspike interval $\langle T
\rangle$ is depicted for a fixed channel noise strength. If the
delay coupling mechanism does not  dominate the spiking, i.e. for
small supra-threshold coupling strengths, the mean interspike
interval exhibits a smooth dependence on the delay time. However for
larger supra-threshold driving strengths, for which the ISIH
exhibits a unimodal structure consisting of a sharp peak, the mean
interspike time varies with the delay time $\tau$ in an almost
piecewise linear fashion,  displaying sharp, triangle-like textures
, cf. Fig.~\ref{fig:mean}. At these sharp peak locations, the number
of spikes that fit according to the intrinsic time into a full time
length given by the delay time, just increases by unity with
increasing delay time $\tau$. The mean interspike interval is
henceforth proportional to the ratio of the delay time and the
number of spikes $n$ fitting into this very delay-time interval;
yielding
\begin{align}
  \label{eq:meaninterspike}
  \langle T \rangle \propto \tau \big/ n \doteq \tau \kappa\, .
\end{align}
In Fig.~\ref{fig:kappa} the behavior of $\kappa$ {\it vs.}  the
delay time $\tau$ is depicted. At multiples of the noise-dependent
intrinsic time scale, such characteristic steps  do occur indeed.

We find that stronger noise intensity mimics a decrease of the
coupling strength. Nearby those typical steps noise is able to
induce newly generated spikes at the end of the intervals reducing
thereby the ratio $\kappa$. Alternatively, the role of noise
decreases the number of spikes towards the left boundary of a new
synchronized region. Therefore the corresponding results at
increased noise strength attain a form that is similar to the case
of  lower delay coupling strength.

A comparable scenario  has been reported in other systems such as
for a noisy  Van der Pol dynamics, see in Refs.
\cite{AMJP,prl04,sch2} near the Hopf bifurcation, for semiconductor
superlattices \cite{sh} and stochastic excitable dynamics
\cite{sch}. We find that noise can induce oscillations with a
well-defined time scale, and the simultaneous  application of a
delay of duration $\tau$ is able to stabilize this orbit.

\subsection{Phase oscillator modeling}
We have seen that the neuron behaves almost like an oscillatory
system whenever the delay coupling rules the dynamics. This brings
to mind a description in terms of a phase oscillator model which we
develop next. Let us assign an intrinsic frequency by $\omega = 2\pi
/ {\cal T}_{\mathrm{int}}$. As pointed out above, the autaptic
feedback leads to a stabilization of the internal rhythmicity as
determined by the model parameters. This in turn results in
synchronized patterns, cf. Fig.~\ref{fig:kappa}.

Therefore we investigate the question whether the observed
synchronization in this studied neuronal model with autapse can be
characterized by a Kuramoto-type model \cite{acebron} of a phase
oscillator including delayed feedback? Put differently, upon
neglecting the details of the shape of the spike trains the
essential information can be reduced to the phase of the oscillatory
neuron.

\begin{figure}
  \centering
  \includegraphics{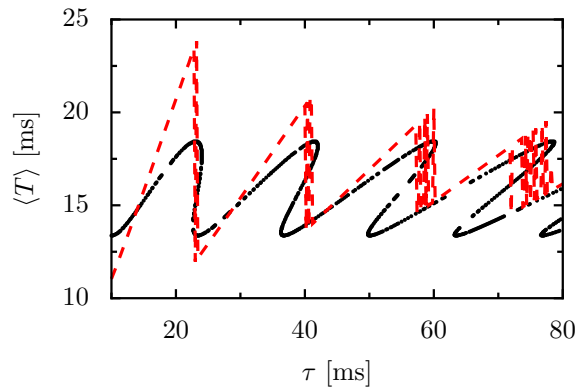}
  \caption{(Color online) Comparison of the stochastic, neuronal
    dynamics Hodgkin-Huxley model including feedback, Eqs.~(\ref{eq:comp_mod}, \ref{eq:gating}, \ref{eq:correlator-a}, \ref{eq:noisecor}) with the dynamics  of the feedback assisted
    Kuramoto-type phase oscillator model, Eq. (\ref{eq:phase-oscillator}). The mean interspike
    interval (red dashed lines) is the result of  the numerical integration
    of the stochastic Hodgkin-Huxley model with delayed feedback;
    there is quantitative good agreement between period of oscillations as obtained from a phase oscillator
    dynamics of the Kuramoto-type delayed feedback model (solid black line).  Regions with multiple solutions in the phase model
    correspond to discontinuous jumps between different types of
    multiplets in the Hodgkin-Huxley model. The parameters for the neuronal
    modeling are chosen as $N_\chem{K}=150$, $N_\chem{Na}=500$ and $\epsilon=0.4
    \un{mS/cm^2}$; those of the phase oscillator are  given by oscillator frequency $\omega = 2\pi /
    15.5\un{ms}$ and feedback strength  $A=1 / 15.5 \un{ms}$. }
  \label{fig:comp}
\end{figure}

Let us next compare our findings with the results obtained from a
pure phase dynamics. Accordingly, the proposed dynamics for the
underlying  phase is modeled as:

\begin{align}
  \label{eq:phase-oscillator}
  \dot{\phi} = \omega - A \sin \left[ \phi(t) - \phi(t-\tau )\right]\, ,
\end{align}
where $\omega$ denotes the angular intrinsic frequency of the
oscillator and $A$ is the delay-coupling strength. Searching for
solutions with fixed frequency $\Omega$, we make the Ansatz:
\begin{align}
  \label{eq:ansatz}
  \phi(t) = \Omega \, t\, .
\end{align}
The resulting, self-consistent equation reads
\cite{schuster1,schuster2,morelli}:
\begin{align}
  \label{eq:Omega}
  \Omega = \omega - A \sin \left[ \Omega \tau \right]\, ,
\end{align}
which was solved numerically. Comparing this phase oscillator
dynamics with our  oscillatory stochastic neuron dynamics exhibiting
an autapse, we make the following parameter identification, i.e.,
$\Omega = 2\pi / \langle T \rangle$ and obtain from an expansion in
small delay time $\tau$ the relation $A = 1 / {\cal
T}_{\mathrm{int}}$. In Fig.~\ref{fig:comp} we present the comparison
between the Hodgin-Huxley modeling and the
the phase oscillator modeling: Even though the quantitative behavior is
different, a good qualitative agreement is detected. The positions
of the spikes where the autapse model jumps between bursts of
different number of multiplets are nicely reproduced by this
simplified phase oscillator modeling. A further  improvement of the
modeling would require a more optimal determination of the
$\tau$-dependent
parameters of this  Kuramoto model with feedback.\\

\section{Conclusion}
\label{sec:conclusion} With this work we have investigated the
effect of delayed feedback on the dynamics of a {\it stochastic}
Hodgkin-Huxley neuron. Using the original Hodgkin-Huxley parameters,
we simulated numerically the stochastic Hodgkin-Huxley model taking
into account the effects of  intrinsic channel noise. A Pyragas-like
delayed feedback mechanism is employed to model the autapse
phenomena, in which a neuron's dendrite back-couples to itself. The
two basic parameters used in our study are the strength of the
autapse-coupling $\epsilon$ and the time-delay $\tau$ resulting from
the finite length of the self-connecting dendrite.

We have found that the neuron dynamics exhibits intriguing new times
scale that stem from the autaptic connection. Since the rest state
of the neuron is always stable, noise or an initial spike is
necessary to create activity, i.e. the spiking dynamics. For a small
number of $\chem{Na}$ and $\chem{K}$ channels, the noise becomes sizable and the
excitory dynamics remains practically unaffected by the delay. In
contrast, smaller noise levels and stronger coupling strengths
induce a different synchronization phenomena between the delay time
and the intrinsic (also noise dependent) time scales. The delay time
and these intrinsic time scales determine how many
spikes will be created and become subsequently locked  during one
delay epoch. We further underline that our exemplary study can
be mimicked  qualitatively in terms of a  reduced description given
by a Kuramoto phase dynamics with built-in feedback.

We have shown, that the delayed feedback mechanism serves as a control option for adjusting
the peaked distribution of interspike intervals, being of importance
 for  memory storage \cite{sto} and stimulus-locked
short-term dynamics in neuronal systems \cite{a1}. One may therefore speculate
whether nature adopted the autapse phenomena for frequencies-filtering
in presence of unavoidable intrinsic channel noise.

\section{Acknowledgments}
This work has been supported by the China Scholarship Council (CSC),
the Volkswagen Foundation (projects I/80424 and I/80425) and the
Bernstein Center for ``Computational Neuroscience'' Berlin.


\begin{thebibliography}{99}
\bibitem{cc1} G. M. S\"{u}el, J. Garc\'ia-Ojalvo, L. M. Liberman and M.
  B. Elowitz, Nature {\bf440}, 545 (2006).
\bibitem{cc2} J. L. Cabrera and J. G. Milton, Phys. Rev. Lett.
  {\bf89}, 158702 (2002).
\bibitem{sh} J. Hizanidis and E. Sch\"oll, Phys. Rev. E {\bf78},
  066205 (2008).
\bibitem{th} T. Erneux and J. Grasman, Phys. Rev. E {\bf78}, 026209
  (2008).
\bibitem{gao} W. Bao, Z. Li, L. Zhou and Z. Gao, Phys. Rev. E {\bf79},
  016214 (2009).
\bibitem{cri-1} S. R\"udiger and L. Schimansky-Geier, J. Theor. Biol.
  {\bf259}, 96 (2009).
\bibitem{cri} C. Masoller, M. C. Torrent and J. Garc\'ia-Ojalvo, Phys.
  Rev. E {\bf78}, 041907 (2008).
\bibitem{gao1} C. Beta, M. Bertram, A. S. Mikhailov, H. H. Rotermund
  and G. Ertl, Phys. Rev. E {\bf67}, 046224 (2003).
\bibitem{sch} T. Prager, H. Lerch, L. Schimansky-Geier and E.
  Sch\"oll, J. Phys. A: Math. Theor. {\bf40}, 11045 (2007).
\bibitem{ku2} G. C. Sethia, J. Kurths and A. Sen, Phys. Lett. A {\bf364},
  227 (2007).
\bibitem{ex} M. A. Arteaga, M. Valencia, M. Sciamanna, H. Thienpont,
  M. L\'opez-Amo and K. Panajotov, Phys. Rev. Lett. {\bf99}, 023903
  (2007).
\bibitem{aut-new1} A. B. Karabelas and D. P. Purpura, Brain Res.
 {\bf200}, 467 (1980).
\bibitem{aut-new2} G. Tam\'{a}s, E. H. Buhl and P. Somogyi, J. Neurosci.{\bf17}, 6352 (1997).
\bibitem{aut-new3} J. M. Bekkers, Curr. Biol. {\bf8}, R52 (1998).
\bibitem{aut-new4} K. Ikeda and J. M. Bekkers, Curr. Biol. {\bf
    16}, R308 (2006).
\bibitem{Loos1972}H. Van Der Loos and E. M. Glaser, Brain Res.
  {\bfseries 48}, 355 (1972).
\bibitem{aut3} J. L\"{u}bke, H. Markram, M. Frotscher and B. Sakmann,
  J. Neurosci. {\bf 16}, 3209 (1996).
\bibitem{Lindner} B. Lindner, J. Garc\'ia-Ojalvo, A. Neiman and L.
  Schimansky-Geier, Phys. Rep. {\bf392}, 321 (2004).
\bibitem{Gammaitoni} L. Gammaitoni, P. H\"anggi, P. Jung and
  F. Marchesoni, Rev. Mod. Phys. {\bfseries 70}, 223 (1998).
\bibitem{hanggi} P. H\"anggi, ChemPhysChem {\bfseries 3}, 285 (2002).
\bibitem{Schmid_EPL} G. Schmid, I. Goychuk and P. H\"anggi, Europhys.
  Lett. {\bf56}, 22 (2001).
\bibitem{Shuai_EPL} P. Jung and J. W. Shuai, Europhys. Lett. {\bf56}, 29
  (2001).
\bibitem{meinhold}L. Meinhold and L. Schimansky-Geier, Phys. Rev E {\bf
    66}, 050901R (2002).
\bibitem{schi1}G. Schmid, I. Goychuk and P. H\"anggi, Phys. Biol.
  {\bf1}, 61 (2004).
\bibitem{schi2} G. Schmid, I. Goychuk and P. H\"anggi, Phys. Biol.
  {\bf3}, 248 (2006).
\bibitem{bur} G. C. Sethia and A. Sen, Phys. Lett. A {\bf359}, 285
  (2006).
\bibitem{sy} A. Pikovsky, M. Rosenblum and J. Kurths, {\it Synchronization:
  A Universal Concept in Nonlinear Sciences} (Cambridge University
  Press, Cambridge, 2003).
\bibitem{sy0} A. M. Lacasta, F. Sagu\'es and J. M. Sancho, Phys. Rev. E
{\bf66}, 045105(R) (2002).
\bibitem{sy1}
L.~Callenbach, P.~ H\"anggi, S. J. Linz, J. A. Freund and L.
~Schimansky-Geier, Phys. Rev. E {\bf 65}, 051110 (2002).
\bibitem{sy2} J. A. Freund, L. Schimansky-Geier and P. H\"anggi,  Chaos {\bf13}, 225 (2003).
\bibitem{noi1} R. F. Fox and Y. Lu, Phys. Rev. E {\bf49}, 3421 (1994).
\bibitem{Ochab} A. Ochab-Marcinek, G. Schmid, I. Goychuk and P.
  H\"anggi, Phys. Rev. E {\bf79}, 011904 (2009).
\bibitem{hh} A. L. Hodgkin and A. F. Huxley, J. Physiol. {\bf117}, 500
  (1952).
\bibitem{white} J. A. White, J. T. Rubinstein and A. R. Kay, Trends
  Neurosci., {\bf23}, 131 (2000).
\bibitem{del} K. Pyragas, Phys. Lett. A {\bf170}, 421 (1992).
\bibitem{tuckwell} H. C. Tuckwell, J. Theoret. Biology {\bf 127}, 427
  (1987).
\bibitem{bm} G. E. P. Box and M. E. Muller, Ann. Math. Stat. {\bf29},
  610 (1958).
\bibitem{textbook} J. Keener and J. Sneyd, Mathematical Physiology
  (Springer-Verlag, New York, 1998)
\bibitem{sch-a} G.~ Schmid, I.~ Goychuk, and P.~ H\"anggi, Physica A
  {\bf325}, 165 (2003).
\bibitem{LNP}
G.~ Schmid, I.~ Goychuk, and P.~ H\"anggi, Lecture Notes in Physics, {\bf 625}, 195 (2003).
\bibitem{ku} C. Zhou and J. Kurths, Chaos {\bf13}, 401 (2003).
\bibitem{AMJP} P. H\"anggi and P.~Riseborough, Am. J. Phys. {\bf 51}, 347
(1983).
\bibitem{prl04} N. B. Janson, A. G. Balanov and E. Sch\"oll, Phys.
  Rev. Lett. {\bf93}, 010601 (2004).
\bibitem{sch2} A. Pototsky and N. Janson, Phys. Rev. E {\bf76}, 056208 (2007).
\bibitem{acebron} J. A. Acebr\'{o}n, L. L. Bonilla, C. J. P. Vicente,
  F. Ritort and R. Spigler, Rev. Mod. Phys. {\bf77}, 137 (2005).
\bibitem{schuster1}H. G. Schuster and P. Wagner, Prog. Theor. Phys. {\bf
    81}, 939 (1989).
\bibitem{schuster2}E. Niebur, H. G. Schuster and D. M. Kammen, Phys. Rev. Lett. {\bf
    67}, 2753 (1991).
\bibitem{morelli}L. G. Morelli, S. Ares, L. Herrgen, C. Schr\"oter, F.
  J\"ulicher and A. C. Oates, HFSP Journal {\bf 3}, 55 (2009).
\bibitem{sto} H. S. Seung, J. Comput. Neurophysiol. {\bf9}, 171
  (2000).
\bibitem{a1} P. A. Tass, Europhys. Lett. {\bf59}, 199 (2002).
\end{thebibliography}
\end {document}